\magnification = \magstep1
\nopagenumbers
\hsize=15.0truecm
\hoffset=3.0truecm
\hoffset=0.5truecm
\vsize=22.5truecm
\voffset=2.5truecm
\voffset=0.0truecm

\output={\plainoutput}
\pretolerance=3000
\tolerance=5000
\hyphenpenalty=10000    
\newdimen\digitwidth
\setbox0=\hbox{\rm0}
\digitwidth=\wd0

\def\footnoterule{\kern-3pt \hrule width \hsize \kern 2.6pt
\vskip 3pt}
\def\cl{\centerline}

\def\vs{\vskip 11pt}

\def\solar{\ifmmode _{\mathord\odot}\else $_{\mathord\odot}$\fi}

\font\ksub=cmsy7
\def\teff{T$_{\kern-0.8pt{\ksub e\kern-1.5pt f\kern-2.8pt f}}$}
%
\vs\vs\vs
\vs\vs\vs
\vs
\vs
\vs
\centerline{THE FORMATION OF LIFE}
\vs\vs
\cl{Robert L. Kurucz}
\vs
\cl{Harvard-Smithsonian Center for Astrophysics}
\vs
\vs
\vs
\vs
\centerline{November 7, 2000}
\eject
\centerline{THE FORMATION OF LIFE}
\vs\vs
\cl{Robert L. Kurucz}
\vs
\cl{Harvard-Smithsonian Center for Astrophysics, 60 Garden St, Cambridge, MA 02138}
\vs\vs
\cl{ABSTRACT}
\vs
     The formation of life is an automatic stage in the consolidation of 
rocky or ``terrestrial" planets.  The organic (=carbonaceous) matter, 
light elements, gases, and water must ``float" toward the surface and the 
heavier metals must sink toward the center.  Random processes in the molecular 
soup that fills microfractures in unmelted crust eventually produce
self-replicating microtubules.  In an appendix I suggest that some primordial
crust remains because there is not enough consolidation energy to melt the 
whole planet.  Energy is lost when iron planetesimals first partially melt 
and then coalesce to form the molten iron planetary core.  Stony planetesimals 
accrete onto the surface of an already consolidated core. 

\vs
\vs
\cl{LIFE}
\vs
     Planetesimals coalesce into a planet and consolidate into a smoothly 
varying solid body with a hot interior.  The organic (=carbonaceous) matter, 
light elements, gases, and water must ``float" toward the surface and the 
heavier metals must sink toward the center.  There is still original crust, 
and perhaps cooled igneous crust from lava flows that is brittle and full 
of fractures, microfractures, and voids down to the level where the internal 
temperature softens the rock and closes microfractures, say ten kilometers.  

The whole ten-kilometer surface layer of this crust fills with organic sepia-colored
soup charged with gases.  The crust acts as a high pressure reaction vessel.  
The microfractures are lined with crystals.  There a thousand different
wild crystals of elements, alloys, minerals, etc. that can provide
catalytic scaffolding on which to build molecular films.  There are millions
of organic molecules that have large representations in the soup.  All the
possible combinations of interactions between the thousand kinds of crystal 
surfaces and the million kinds of organic molecules are automatically tried 
by the soup.  Longish molecules that are hydrophilic on one end and hydrophobic
on the other and have a ``diameter" that is a multiple of the crystal spacing
can form a monolayer with the molecules aligned side by side with the 
hydrophylic ends weakly bonding to the crystal surface and with the hydrophobic
ends quasi-replicating the crystal surface as in Figure 1.  If the crystal 
structure is not rectilinear, the structure of the film is not rectilinear.  
The edges where the film is growing can be stepped or ragged.  

     All the possible combinations of interactions between the new monolayer 
surfaces and the million kinds of organic molecules are automatically tried by 
the soup.  Some molecules work and attach to the hydrophobic ends and to each 
other.  They produce a bilayer that is stable and whose surface quasi-replicates 
the crystal.  

     The microfractures are not a quiet environment. There are accidents, 
earthquakes, impacts, sudden flows of soup, etc that dislodge the bilayers from 
the crystal faces.  Free in the soup, the bilayers curl with the hydrophobic 
face on the outside.  If opposite edges are straight, the edges can curl together
and bind into microtubules.  They become bilayer cylinders with hydrophilic outer surfaces 
and hydrophobic interior surfaces.  Soup flows through and over the microtubules.
Some of the microtubules are made with combinations of molecular layers that are 
self-catalyzing and self-propagating.  The bilayer edges continue to bind more 
molecules to both layers at both ends of the cylinder.  The microtubules grow 
longer and longer.  Eventually they mechanically break and each piece becomes 
a growing microtubule.  Each microtubule created this way from a crystal template
is a new self-replicating life form.  The crust of the planet 
fills with microtubules.  There are thousands of different kinds, each in a range 
of diameters that were accidentally determined.  Life is automatically produced 
thousands of times.  

  In Figure 1, the g molecule descenders all point the same way.  If the original 
bilayer was made from chiral molecules, the microtubule descendents will all have 
the same chirality.

Each microtubule is a ``catalytic converter" because the inside and the outside 
are a quasi-replicas of a crystal.  From the soup, on their inner and outer 
surfaces, they form dimers, trimers, polymers, polymers of dimers, etc.  The 
microtubules can hold two molecules stationary while they interact with each 
other.  They can ``crack" or disassemble large molecules by holding one end
so the rest can be attacked by the soup.  The microtubules can make molecules 
and polymers that do not release but are bound to the wall.  An elastic polymer 
would allow the microtubule to move in response to stimuli.  If an end hit 
something hot the polymer would contract and squirt soup out the end in a jet.  
Or peristalsis would be possible to pump the soup through the tube to increase 
reaction rates.  The microtubules can bind to each other.

There are, say,  10$^{12}$ generations in the first billion years.  Since the 
original microtubules were made by random processes, they many not be optimal.  
At the growing ends of the microtubules, a molecular substitution can be made if 
some other molecules will bind into that space.  The random processes also test 
all possible isotopomeric substitutions.  The new molecule can be a 
``dud" that truncates growth in that position.  The microtubule grows narrower. 
If it happens a few times the whole end of the microtubule becomes closed.
The new substitute molecules may bind more strongly or less, faster or slower, 
strengthen or weaken the backbone.  Changes that make the microtubule more 
fit will be adopted.  If the change makes a whole row grow faster,
it can stick out of the end of the microtubule as a microrod.  Eventually the rod 
breaks off the microtubule leaving a free-floating microrod.  Microrods have 
the same replicating and catalyzing capablity as the microtubules.  The microrods 
spread throughout the crust.   Microrods can act as carriers that hold small 
molecules in position for other reactions.

The microtubules also make substitution errors in building polymers that sometimes
are an improvement.

All of this is independent of conditions on the surface of the crust.

Once there is cold fractured lava on the surface of the planet and the soup 
has soaked through it, the primordial crust is no longer needed.        

Life becomes more and more complex and may eventually spread from the interior
of the crust to the surface as well.

\vs
\vs

\cl{APPENDIX: IRON CORES}
\vs

There must be some sort of galactic or protostellar magnetic field in a
protoplanetary disk.  When iron condensates form they are magnetized by the
field.  The iron particles attract each other and form larger and larger
agglomerations that eventually collapse under their own weight and heat 
themselves.  As the agglomerations grow the iron melts in the densest regions.

Meteorites are pieces of asteroids.  Asteroids are planetesimals that failed
to form a planet because there were not enough of them.  Some meteorites are solid
iron, some are stony with iron, and some are stony.  Some of the stony
meteorites are carbonaceous chondrites that have organic inclusions that have 
never been baked.  The asteroids must have the same properties as the 
meteorites.  Therefore molten iron can be formed in a body the size of an
asteroid.  Planetesimals rotate.  If they have a molten iron core they generate
a magnetic field.  They attract other planetesimals that have iron.  They attract
iron condensates from the disk.  They tend to form a planetary core before the
stony planetesimals are agglommerated.  The heating from the 
energy of consolidation mostly takes place before the planet is formed so that
much of it can radiate away and not be trapped in the planet.  Thus the planet
does not heat internally as much as has been thought in the past.  Heating from
radioactive decay is not affected and continues to be important.  The total heating 
is not strong enough to melt the whole stony surface of the planet.  The crust 
has much organic (= carbonaceous) material.

\vfill
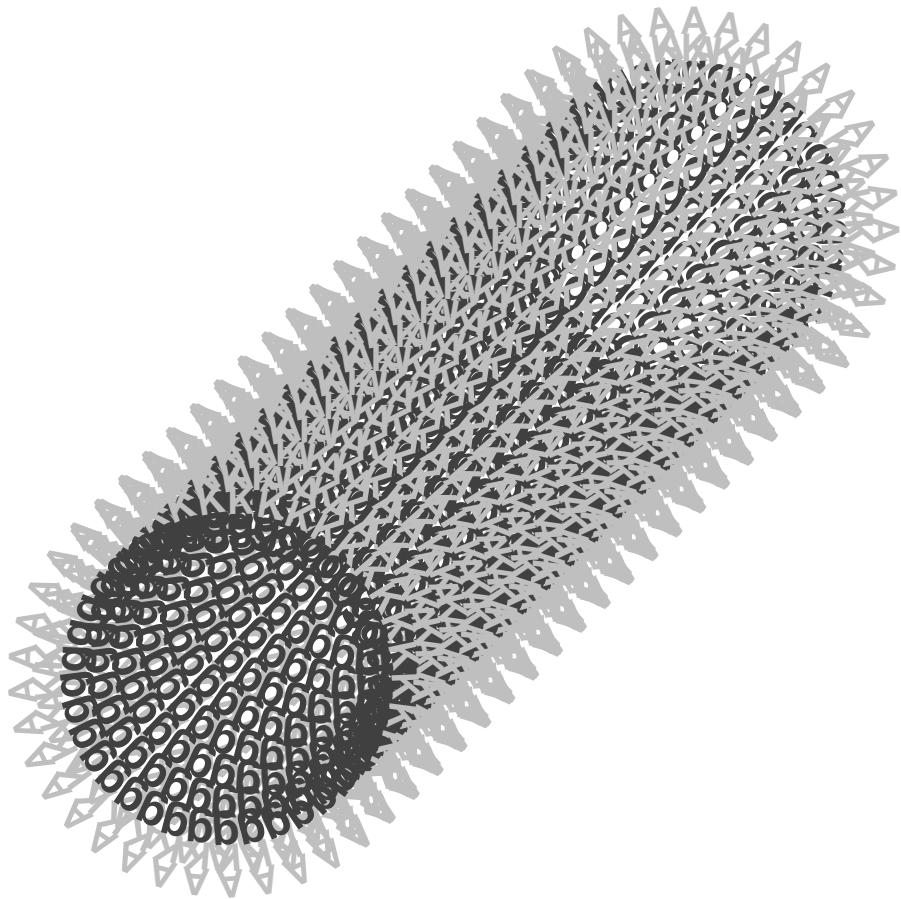
\end